\documentclass[twocolumn,prl,preprintnumbers,footinbib]{revtex4}

\usepackage{graphicx}
\usepackage{dcolumn}
\usepackage{bm}
\usepackage{hyperref}

\begin{document}

\title{Theory of fast optical spin rotation in a quantum dot based on geometric phases and trapped states}

\author{Sophia E. Economou}
\author{T. L. Reinecke}

\affiliation{Naval Research Laboratory, Washington, DC 20375, USA}
\date{\today}

\begin{abstract}
A method is proposed for the optical rotation of the spin of an
electron in a quantum dot using excited trion states to implement
operations up to two orders of magnitude faster than those of most
existing proposals. Key ingredients are the geometric phase induced
by 2$\pi$ hyperbolic secant pulses, use of coherently trapped states
and use of naturally dark states. Our proposal covers a wide variety
of quantum dots by addressing different parameter regimes. In one
case the treatment provides an exact solution to the three-level
system. Numerical simulations with typical parameters for InAs
self-assembled quantum dots, including their dissipative dynamics,
give fidelities of the operations in excess of 99\%.
\end{abstract}

\maketitle

All-optically controlled spins in quantum dots (QDs) provide an
attractive proposal for quantum information processing (QIP)
\cite{imamoglu}. This proposal combines the merits of solid state,
such as integrability with existing semiconductor technology, with
the speed and high degree of control of lasers. Among the
requirements for QIP \cite{divincenzo_criteria}, optical preparation
\cite{gurudev} and measurement \cite{cortez_prl,comb_readout} of the
spin qubit have been achieved, and the spin coherence time has been
measured to be at least 3 $\mu$s \cite{greilich_science}. At the
heart of this QIP approach are optical spin rotations that
constitute the quantum gates by which information is manipulated.
This key step toward the development of optical QIP with spins in
QDs has not yet been demonstrated experimentally, although several
proposals have been given
\cite{pochung,emary_rotation,kisrenzoni,troiani_gates,economouprb06}.
Most of the latter employ the adiabatic approximation either through
adiabatic elimination of an excited state
\cite{pochung,emary_rotation} or through Stimulated Raman Adiabatic
Passage \cite{kisrenzoni,troiani_gates}, and thus they rely on
adiabatic pulses that limit the speed of the gates. A method for
fast optical spin rotations about a single axis (the optical axis)
was proposed \cite{economouprb06}, but rotations about arbitrary
axes are required for QIP.

Here we present a proposal for achieving arbitrary coherent spin
rotations without making use of an adiabatic approximation. We show
that for realistic QD systems our method attains high (up to
99.99\%) fidelity gates operating up to two orders of magnitude
faster than other proposals. Our approach involves naturally dark or
coherently trapped states (depending on the system parameters
involved) combined with novel use of the geometric phase analysis
\cite{economouprb06} for the hyperbolic secant pulses of Rosen and
Zener \cite{RZ}.

These sech pulses have the remarkable property of rendering the
resulting time dependent Schr\"odinger equation for a two-level
system analytically solvable. A pulse area can be defined for them
independent of the detuning, an asset that led to the celebrated
phenomenon of self-induced transparency \cite{SIT} for 2$\pi$
pulses. Such pulses propagate virtually unattenuated through
resonant optical media, returning the individual two-level systems
to their initial states. This is a highly attractive feature for the
current QIP proposal because the information must be returned and
stored in the spin subspace after it is manipulated. A less explored
effect is that after the passage of the 2$\pi$ sech pulse the state
also acquires a geometric phase, for which we obtained an analytic
expression \cite{economouprb06}. This phase is global (an overall
phase to the quantum state) for a two-level system. It becomes
crucial when a third state that does not couple to the laser is
present, and it is one of the key ingredients of the present
proposal.

\begin{figure}
\begin{center}
\includegraphics[height=3.4cm,width=5.8cm ]{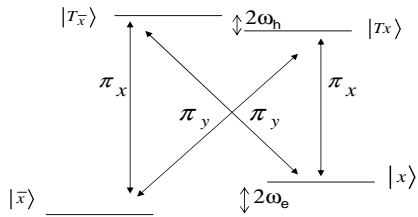}
\end{center}
\caption{Energy levels of the four-state system, which is comprised
of the two electron spin eigenstates of $\sigma_x$,
\{$|x\rangle,|\bar{x}\rangle$\} and the two trion spin states.
Linearly polarized light $x$ ($y$) denoted by $\pi_x$ ($\pi_y$)
induces only the indicated transitions.} \label{4levels}
\end{figure}

 The energy levels and selection rules of the spin system in our approach are
shown in Fig. \ref{4levels}. An external magnetic field B along the
in-plane ($x$) direction defines the quantization axis for the
electron spin; the optical axis, which coincides with the growth
axis of the QD, defines the $z$ direction. The intermediate states
are the so-called trion states, which are bound states of an
electron and an exciton, the latter created in the QD by the laser.
The angular momentum of the single electron levels comes from its
spin. The hole has a total angular momentum of $3/2$, and the
relevant excited states are the $m_j=\pm 3/2$ (`heavy hole') trion
states quantized along $z$. In typical QDs, the $m_j=\pm 1/2$
(`light hole') states are separated by more than 20 meV from the
$m_j=\pm 3/2$ states due to confinement, and they are not included
here. For a magnetic field along $x$ the trion energy eigenstates,
$|T_x\rangle,~|T_{\bar{x}}\rangle$, are linear combinations of the
heavy hole trions. Fig. \ref{4levels} shows the optical selection
rules: the transitions $|x\rangle \rightarrow |T_x\rangle$ and
$|\bar{x}\rangle \rightarrow |T_{\bar{x}}\rangle$ ($|x\rangle
\rightarrow |T_{\bar{x}}\rangle$ and $|\bar{x}\rangle \rightarrow
|T_x\rangle$) are coupled by linearly polarized light $\pi_x$
($\pi_y$). The Zeeman splittings for the electron and the trion are
$2 \omega_e$ and $2 \omega_h$ respectively. The trion splitting is
determined by the hole \textsl{g} factor because the electrons are
in the same orbital state of the QD (a spin singlet) and do not
contribute to the total trion spin.

Our proposal is based on the observation that from the four-level
system of Fig. \ref{4levels}, different two-level systems can be
selected by an appropriate choice of laser polarization without the
need for frequency selectivity. In our approach all four states
participate in the rotation scheme. For each of the decoupled
two-level systems, a 2$\pi$ sech pulse will be used to induce the
desired phase in the ground state. It is well known that given
arbitrary rotations by two axes any rotation can be implemented as a
composite rotation. Here we design rotations about the $z$ and the
$x$ axes by arbitrary angles and compose general rotations from
them.

 \emph{(a) Rotations about $z$}- We showed earlier
\cite{economouprb06} that broadband circularly polarized pulses
(here determined by the bandwidth $\beta_z \gg 2(\omega_e +
\omega_h)$) allow the four-level system in Fig. \ref{4levels} to be
treated as a two-level system (plus two uncoupled levels): During
the ultrafast pulse the spin and trion precessions are `frozen'. The
phase gates of Ref. \cite{economouprb06} are used for $z$ rotations.
The angle of rotation is
\begin{eqnarray}
\phi_z = 2 \arctan(\beta_z/\Delta),\label{phiz}
\end{eqnarray}
where $\Delta$ is the detuning. The fidelity and purity of the $z$
operations both increase as the pulses become faster
\cite{economouprb06}. For a subpicosecond pulse, we found the
fidelity to be as high as 99.99\%.
\begin{figure}
\begin{center}
\includegraphics[height=4.8cm,width=5.6cm ]{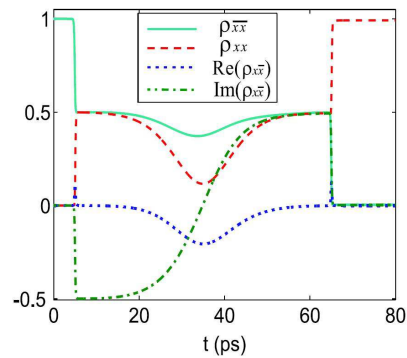}%{ME}{MEs4new} width 5.5
\end{center}
\caption{(Color online) The matrix elements of the spin density
matrix, $\rho$, for the rotation $R_y(\pi)$. The red (dashed) line
is $\rho_{xx}$, the cyan (solid) line is $\rho_{\bar{x}\bar{x}}$,
the blue (dotted) line is Re$(\rho_{x\bar{x}})$, and the green
(dashed-dotted) line is Im$(\rho_{x\bar{x}})$, and $\rho$ is the
density matrix in the interaction picture (free spin precession is
removed). The duration of this composite gate is about 60 ps.}
\label{ypi}
\end{figure}
\begin{figure}
\begin{center}
\includegraphics[height=4.2cm,width=5.1cm ]{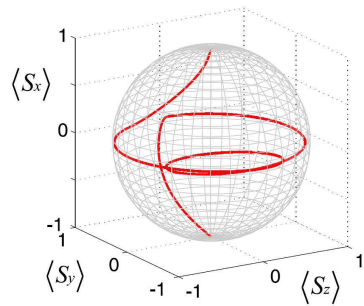}%{bs4new}
\end{center}
\caption{(Color online) Bloch sphere representation of the composite
spin rotation $R_y(\pi)$ acting on the initial spin state
$|\bar{x}\rangle$, i.e., $\langle S_x\rangle=-1$.} \label{ypiBS}
\end{figure}

 \emph{ (b) Rotations about $x$}- As illustrated in Fig. \ref{4levels}
by choosing linearly polarized light $\pi_x$, we reduce the
four-level system to two independent two-level systems,
\{$|x\rangle, |T_x\rangle$\} and \{$|\bar x\rangle,
|T_{\bar{x}}\rangle$\}. A $\pi_x$ linearly polarized 2$\pi$ sech
pulse with bandwidth $\beta_x$ is used here, which we allow to act
on both transitions and which is detuned by $\Delta_1$ ($\Delta_2$)
for transition $|x\rangle \leftrightarrow |T_x\rangle$ ($|\bar
x\rangle \leftrightarrow|T_{\bar{x}}\rangle$). Each two-level system
acquires a phase (Eq. \ref{phiz}) determined by its detuning. The
relative phase determines the angle of rotation about the $x$ axis,
$\phi_x=\phi_1-\phi_2$, which is
\begin{eqnarray}
\phi_x = 2\arctan{\frac{\beta_x (\Delta_1-\Delta_2)}{\Delta_1
\Delta_2 + \beta_x^2}} \label{phix}
\end{eqnarray}
in terms of pulse and system parameters. From Eq. (\ref{phix}) it is
clear that a requirement for a $\pi$ rotation is $\Delta_1\Delta_2 <
0$. This can be understood intuitively: If the laser is detuned
positively (or negatively) for \emph{both} transitions, then the
phases will have the same sign, so there is an upper limit to the
obtainable angle of rotation. Centering the laser frequency between
the two transitions gives rotation angles up to $\pi$ provided that
$\beta_x$ is chosen properly. Therefore for a $\pi$ rotation there
is an upper bound to the bandwidth $\beta_x$, which makes rotations
about $x$ slower than those about $z$.

 \emph{(c) Rotations about other axes}- Finally, by combining the
above rotations about $x$ and $z$ we can implement any rotation. For
example, rotations about $y$ can be realized by $R_y(\phi) =
R_z^\dag(\pi/2) R_x(\phi)R_z(\pi/2)$ or alternatively by sandwiching
a $z$ rotation between two $x$ rotations.

 The imperfections in these gates come predominately from trion decay
and spin precession during $z$ rotations. These effects are taken
into account in our calculations of the fidelity by numerical
solution of the Liouville equation for the density matrix. The
fidelity is a measure of how well the gate is implemented and is
defined as $\mathcal{F}(U)=\overline{|\langle\Psi|{U}^{\dag}
U_{id}|\Psi\rangle|^2}$, where $U_{id}$ is the target operation, $U$
is the actual operation, and the average is taken over all input
spin states \cite{bowdrey_fidelity}. The purity of an operation is
$\mathcal{P}=\overline{\text{Tr}\rho^2}$, where $\rho$ is the spin
density matrix after the rotation.

 We have made calculations of the fidelity and purity of gates
using parameters appropriate for realistic self-assembled InAs QDs:
$\omega_e=0.05~\text{meV}, ~\omega_h=0.033~\text{meV}$
(corresponding to B$\sim$8 T), $\beta_x=0.073$ meV for
$\pi$-rotations, $\beta_x=0.113$ meV for $\pi/2$-rotations,
$\beta_z=4$ meV, and trion lifetime $\tau_t = 900$ ps \cite{ware}.
Typical gate fidelities, listed for some rotations in Table
\ref{table2} along with the purities, are on the order of 99.5\%.

 In Fig. \ref{ypi} we show the corresponding elements of the spin density matrix
for rotation $R_y(\pi)= R_z^\dag(\pi/2) R_x(\pi)R_z(\pi/2)$ acting
on the initial spin state $|\bar{x}\rangle$. The $z$ rotations are
so fast that on this time scale they appear as vertical steps at
$t\simeq 5$ ps and $t\simeq 65$ ps. The total duration of the gate
is also fast, about 60 ps. The corresponding spin vector,
constructed as $\vec{S}=\text{Tr}(\vec{\sigma} \rho_s)$, is shown in
a Bloch sphere in Fig. \ref{ypiBS}. Here $\vec\sigma$ is the spin
operator and $\rho_s$ is the spin density matrix in the
Schr\"odinger picture.

\begin{table}[htp]
\caption{Fidelity and purity of selected rotations of spin for InAs
QD parameters.
 \label{table2}}
\begin{tabular}{|c|c|r|}
    \hline
$R_n(\phi)$ &  Fidelity    &   Purity   \\
    \hline
$R_x(\pi/2)$   & 99.49\% & 98.89\%  \\
$R_z(\pi/2)$     & 99.99\% & 99.97\%  \\
$R_y(\pi)$   & 99.28\% & 98.57\%  \\
$R_y(\pi/2)$   & 99.45\% & 98.8\%  \\
    \hline
\end{tabular}
\end{table}

 Our gate scheme is especially attractive for experimental demonstration of
optical spin rotations in these systems because it uses only simple,
nonadiabatic sech pulses. It does not require phase locking of the
lasers in this parameter regime, which is a significant experimental
simplification. It requires the ability to perform Rabi oscillations
between the spin and the trion, which has been demonstrated recently
for InAs QDs \cite{greilich}, and thus it is experimentally
accessible with state of the art technology. The only approximation
in our method is that the spin is considered to be `frozen' during
the pulse for $z$ rotations, i.e., $\beta_z \gg 2 (\omega_e +
\omega_h)$. This requirement can be satisfied for the widely used
ultrafast lasers and the current generation of InAs QDs.

 For QDs with larger Zeeman splittings the pulse durations
discussed above can become too short for practical purposes.
Examples of such systems include some CdSe QDs. For such systems our
approach can be modified. We utilize the phenomenon of Coherent
Population Trapping (CPT), in which a superposition of two levels,
each level having nonzero dipole coupling to an excited state, is
dark to an (appropriately chosen) coherent combination of laser
pulses. This well-known phenomenon in the optics of atoms \cite{cpt}
has recently been demonstrated in a semiconductor \cite{kaimei} for
a similar system to the one we study. Here we use frequency
selectivity to isolate a three-level system from the four levels of
Fig. \ref{4levels}. We do this by choosing narrowband pulses, $\beta
\ll 2(\omega_e - \omega_h)$, where $\beta$ is the bandwidth. For
concreteness we spectrally focus the pulses to pick out the
$\Lambda$ system consisting of $\{|x\rangle, |\bar x \rangle,
|T_x\rangle\}$. The two transitions are addressed separately by
polarization selectivity. The total laser field is
\begin{eqnarray}
\vec{E} = E_x~f_x(t)~e^{i \omega_x t} \hat{x} + e^{i\alpha}
E_y~f_y(t) ~e^{i \omega_y t} \hat{y} + c.c.
\end{eqnarray}

 To create a coherently trapped state the two pulses should have the
same detuning (two-photon resonance), i.e., $\omega_{x,T_x}-\omega_x
= \omega_{\bar{x},T_x} - \omega_y \equiv \Delta$, and the same
temporal envelope $ f_x(t) = f_y(t) \equiv f(t)$. Then the new spin
states, $|B\rangle$ and $|D\rangle$ (bright and dark respectively),
are related to the basis states \{$|x\rangle, |\bar{x}\rangle$\}
through the unitary transformation
\begin{eqnarray}
\mathcal{T}=
\left[\begin{array}{cc} \cos\vartheta & -e^{i\alpha}\sin\vartheta  \\
e^{-i\alpha}\sin\vartheta  & \cos\vartheta
\end{array}\right].
\label{transf}
\end{eqnarray}
Here $\vartheta$ is defined by
\begin{eqnarray}
\tan\vartheta = E_y / E_x,
\end{eqnarray}
and the matrix element $V_{B,T_x}$ between $|B\rangle$ and
$|T_x\rangle$ is
\begin{eqnarray}
V_{B,T_x} = \Omega_o~f(t)~e^{i \Delta t},
\end{eqnarray}
where $\Omega_o = \sqrt{\Omega_x^2 + \Omega_y^2}$
\cite{eberlykozlov} and $\Omega_x$ ($\Omega_y$) is the Rabi
frequency of the transition with polarization $\pi_{x}$ ($\pi_{y}$).
Now we choose the envelope to be $f(t) = \text{sech}(\beta t)$. We
require the population to return to the spin subspace after the
passage of the pulse, i.e., we require the total pulse acting on the
bright state to be a 2$\pi$ pulse, which gives
\begin{eqnarray}
\Omega_o = \beta.
\end{eqnarray}

 We still have freedom in choosing the bandwidth $\beta$ and the
detuning $\Delta$. The bandwidth will be constrained by the
requirement of frequency selectivity, $\beta \ll 2(
\omega_e-\omega_h)$. After the passage of the pulse the  bright
state has picked up a phase $\phi$ relative to the dark state given
by \cite{economouprb06}
\begin{eqnarray}
\phi = 2 \arctan\left(\frac\beta\Delta\right).\label{phi}
\end{eqnarray}
This phase is the angle by which any spin state not parallel to
$|B\rangle$ (or, equivalently, $|D\rangle$) is rotated. The axis of
rotation is that defined by $|B\rangle$. By varying the laser
parameters, $\vartheta$ and $\alpha$, we vary the composition of the
bright/dark basis, i.e., vary the axis of rotation. Thus we have
designed an arbitrary spin rotation
\begin{eqnarray}
R_{n}(\phi) = e^{-i\phi\hat{n}\cdot\vec{\sigma }/2},
\end{eqnarray}
where $\vec\sigma$ is the spin operator. $\phi$ is given by Eq.
(\ref{phi}) and the axis of rotation by
\begin{eqnarray}
\hat{n}=(\cos\vartheta,\sin\vartheta\sin\alpha,\sin\vartheta\cos\alpha).
\end{eqnarray}
This axis is the same as that of the Kis and Renzoni scheme
\cite{kisrenzoni} where CPT also is used. However, in their approach
an auxiliary lower level is required for the population to be stored
during the gate. This extra requirement limits the applicability of
their scheme to physical systems where such an extra level is
available. In our case, no auxiliary levels are needed, and the same
pulse creates the trapped state and induces the geometric phase.
\begin{figure}
\begin{center}
\includegraphics[height=4.6cm,width=5cm ]{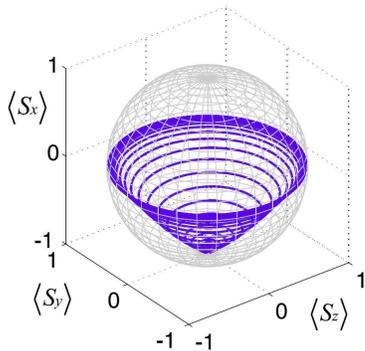}%{bs1}
\end{center}
\caption{(Color online) Bloch sphere representation of the spin
rotation $R_z(\pi/2)$ acting on the initial spin state
$|\bar{x}\rangle$, i.e., $\langle S_x\rangle=-1$.} \label{zpiO2BS}
\end{figure}
\begin{figure}
\begin{center}
\includegraphics[height=4.8cm,width=5.6cm ]{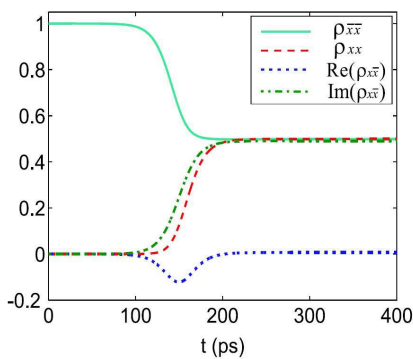}%{rot}
\end{center}
\caption{(Color online) The matrix elements of the density matrix
corresponding to $R_z(\pi/2)$, as in Fig. \ref{zpiO2BS}. The red
(dashed) line is $\rho_{xx}$, the cyan (solid) line is
$\rho_{\bar{x}\bar{x}}$, the blue (dotted) line is
Re$(\rho_{x\bar{x}})$, the green (dashed-dotted) line is
Im$(\rho_{x\bar{x}})$. $\rho$ is the density matrix in the
interaction picture. The duration of this gate is about 100 ps.}
\label{zpio2}
\end{figure}

The main sources of dissipation in the CPT-based scheme are
non-resonant transitions to the higher trion state
$|T_{\bar{x}}\rangle$ and spontaneous emission of the trion. To
determine their effects on the gates we have made numerical
simulations of the fidelity. We choose parameters appropriate for
CdSe nanocrystals as an example of a physical system with relatively
large Zeeman splittings: $\omega_e=0.4$ meV, $\omega_h=0.05$ meV
(corresponding to B$\sim 8$ T), $\beta\sim0.025$ meV, and trion
lifetime $\tau_t = 580$ ps \cite{seufert_prl}. We find that for gate
durations of about 100 ps typical fidelities range between 98\% and
99\%, and they are shown in Table \ref{table1}. A Bloch sphere
representation of the spin rotation $R_z(\pi/2)$ is shown in Fig.
\ref{zpiO2BS}, and the corresponding spin density matrix elements
are plotted as functions of time in Fig. \ref{zpio2}. This figure
shows the total duration of the gate and also shows that the
population $(\rho_{xx}+\rho_{\bar{x}\bar{x}})$ remains mostly in the
spin subspace during the greatest part of the gate.

 The above CPT scheme is an exact analytical solution for a true
three-level $\Lambda$ system. We note that as the Zeeman splitting
of either the electron or the trion gets larger the system more
closely approaches a $\Lambda$ system. In that case fast pulses can
be used to beat the excited state decay time, and the fidelities
then approach 100\%. This rotation scheme may also be used to
implement gates in QIP with other quantum systems such as ions or
atoms where the excited states are typically longer lived than those
of QDs.

\begin{table}[htp]
\caption{Fidelity and purity of selected rotations  \label{table1}
for CdSe QD parameters.}
\begin{tabular}{|c|c|r|}
    \hline
$R_n(\phi)$ &  Fidelity    &   Purity   \\
    \hline
$R_z(\pi/2)$   & 98.84\% & 97.8\%  \\
$R_z(\pi)$     & 97.56\% & 95.34\%  \\
$R_x(\pi/2)$   & 98.42\% & 97\%  \\
$R_{xz}(\pi/2)$   & 99.2\% & 98.78\%  \\
%$R_z(\pi/2)$   & 99.33\% & 98.74\%  \\
%$R_z(\pi)$     & 98.56\% & 97.2\%  \\
%$R_x(\pi/2)$   & 99\% & 98.26\%  \\
%$R_{xz}(\pi/2)$   & 99.4\% & 99.28\%  \\
    \hline
\end{tabular}
\end{table}

 In conclusion, we have developed a method of realizing the key step
of optical spin rotations in realistic QDs based on the geometric
phases induced by hyperbolic secant pulses of area $2\pi$. They are
used with either dark or coherently trapped states, depending on the
parameters of the QDs. This scheme should be accessible with current
experimental capabilities, and it yields fast (1ps-100 ps) rotations
because it does not employ the adiabatic approximation. The
fidelities of these rotations for typical QD systems are high, up to
99.99\%. Our method provides an exact solution to rotations in a
three-level $\Lambda$-type system and may prove useful to other
areas of physics that involve interaction of coherent radiation with
quantum systems \footnote{An additional advantage of our scheme is
that the use of 2$\pi$ sech pulses may allow for potential
manipulation by the same pulse of a logical qubit comprised by
several physical qubits located along the optical path. Then, as in
self-induced transparency, the different qubits would interact with
a single reshaped pulse. We note that numerical simulations
\cite{panzarini} imply that self-induced transparency is possible in
an array of QDs.}.

 This work is supported by the US Office of Naval Research. One of
us (S.E.E.) is an NRC/NRL Research Associate.

%\bibliography{rotation}

\end{document}